\documentclass[11pt]{article}

\def\np#1#2#3{{\it Nucl. Phys.} {\bf B#1} (#2), #3}
\def\pl#1#2#3{{\it Phys. Lett. }{\bf B#1} (#2), #3}
\def\physrev#1#2#3{{\it Phys. Rev.} {\bf D#1} (#2), #3}
\def\jhep#1#2#3{{\it JHEP} {\bf #1} (#2), #3}
\def\hepth#1{hep-th/#1}

\def\Ref#1{(\ref{#1})}

\newcommand{\be}{\begin{equation}}
\newcommand{\ee}{\end{equation}}
\newcommand{\beq}{\begin{eqnarray}}
\newcommand{\eeq}{\end{eqnarray}}

\def\Rb{{\rm \bf R}}
\def\Cb{{\rm \bf C}}

\def\({\left(}
\def\){\right)}
\def\[{\left[}
\def\]{\right]}
\def\rangl{\right\rangle}
\def\langl{\left\langle}
\def\p{\partial}
\def\hf{{1\over 2}}

\def\CF {{\cal F}}

\def\CP {{\cal P}}
\def\CR {{\cal R}}

\def\xp{x_{_{+}}}
\def\xm{x_{_{-}}}
\def\xpm{x_{_{\pm}}}

\def\mub{\mu_{_B}}
\def\mul{\mu_{_L}}
\def\laml{\lambda_{_L}}
\def\mubc{\mu_{_{B,c=1}}}
\def\muc{\mu_{_{c=1}}}

\def\oR{{1\over R}}
\def\ZFZZ{Z^{_{FZZ}}}
\def\ZZZ{Z^{_{ZZ}}}
\def\ZD{Z^{c=1}_{\rm Dir}}

\def\frc#1#2{{\textstyle{#1\over#2}}}
   
\begin{document}



\title{Complex curves and non-perturbative effects in $c=1$ string 
theory\footnote{Contribution to the proceedings of RTN Workshop ``The 
quantum structure of spacetime and the geometric nature of fundamental interactions"
in Kolymbary, Greece, 5-10 September 2004.}}
\author{Sergei Alexandrov\footnote{e-mail: {\sf S.Alexandrov@phys.uu.nl}}} 
\date{}

\maketitle 

\vspace{-1cm} 
\begin{center}
{\it Institute for Theoretical Physics \& Spinoza Institute, \\
Utrecht University, Postbus 80.195, 3508 TD Utrecht, The Netherlands}
\end{center}

\vspace{0.5cm}
\begin{abstract}
We investigate a complex curve in the $c=1$ string theory which provides a geometric 
interpretation for different kinds of D-branes.
The curve is constructed for a theory perturbed by a tachyon potential
using its matrix model formulation. The perturbation removes the degeneracy of the
non-perturbed curve and allows to identify its singularities with ZZ branes.
Also, using the constructed curve, we find non-perturbative corrections 
to the free energy and elucidate their CFT origin. 
\end{abstract}


\vspace{0.5cm}

\section{Inrtoduction}

The $c=1$ string theory represents an interesting laboratory to study many phenomena
inherent to critical string theory (for reviews, see \cite{KLEBANOV,Moore,SAthese}). 
As it was realized recently, it possesses
different kinds of D-branes and its formulation in terms of matrix quantum
mechanics (MQM) can be seen as a kind of open/closed string duality 
\cite{McGreevyKB,MARTINEC,KlebanovKM}.
At the same time, this theory is exactly solvable. Due to this it allows 
to test the ideas and to discover interesting structures which it is difficult to see 
in more complicated cases.

One of such structures which, although known long ago, appeared in the recent
analysis of D-branes in non-critical string theories is a complex curve
capturing different aspects of the theory.
In non-critical strings it is associated with any closed string background
and incorporates information about both 
closed and open string amplitudes.
  
However, such a complex curve was constructed and interpreted in terms of D-branes
only in the case of $c<1$ string theories \cite{Seib,KK}. The $c=1$ limit of this construction 
turns out to be degenerate. As a result, several conclusions achieved for $c<1$
were not evident in the $c=1$ case. 
In particular, the D-brane content was not understood
although some predictions were made from the matrix model analysis \cite{SAmn}.

In this paper we review how one can overcome these difficulties considering the $c=1$ string 
theory perturbed by a tachyon potential \cite{SAcurve}. Since the CFT technique is not enough 
powerful to carry out calculations in such situation, we use the MQM formulation 
to solve the theory and to construct the associated complex curve.
Besides, using the resulting curve, we find non-perturbative corrections
to the free energy which are related to D-instantons in two-dimensional string theory. 
Finally, we interpret the matrix model results in the CFT terms. In particular, we show
the relation of the CFT complex curve to the matrix model one and identify the origin
of the non-perturbative contributions with a subset of ZZ branes.

\section{D-branes and complex curves in minimal string theories}

\subsection{D-branes in Liouville theory}

The recent progress in understanding of non-perturbative effects in non-critical 
string theories is related with the discovery of conformally invariant boundary
conditions (D-branes) in Liouville theory \cite{FZZbrane,ZZ}.
The latter is defined by the following action:
\be
S_L= \int_{\Sigma} {d^2 z\over 4\pi}\sqrt{g} \left( (\p\phi )^2 + Q\hat R\phi
+4\pi\mul e^{2b \phi} \right).
\label{LIOU}
\ee
The central charge of this CFT is given by
\be
c_L=1+6Q^2 
\label{cliouv}
\ee
and the parameter $b$ is related to Q via the relation
\be
Q=b+{1/ b}.
\label{bQ}
\ee
In string theory these parameters are determined by the requirement that the
total central charge of matter and the Liouville field is equal to $26$.
If matter is represented by a minimal $(p,q)$ model with the central charge
$c_{p,q}=1-6{(p-q)^2\over pq}$, the relation \Ref{cliouv} implies that
$b=\sqrt{p/q}$, whereas the coupling to the $c=1$ matter corresponds to the limit $b\to 1$.

The Liouville theory possesses two types of D-branes, or boundary conditions of 
Neumann and Dirichlet type.
The former, the so called FZZ branes, correspond to the following additional boundary term 
\be
S_{\rm bnd}=\int_{\p\Sigma} d\xi \,g^{1/4} \( {Q\hat K\over 2\pi }\phi +\mub e^{b\phi}\),
\ee
and they are parameterized by the boundary cosmological constant $\mub$. In fact, 
it is more convenient to work with another parameter $s$
which is related to $\mub$ through the following relation
\be
{\mub}=\sqrt{\mul\over \sin(\pi b^2)}\,\cosh(\pi b s) .
\label{fzzsmub}
\ee
At the quantum level, the FZZ brane can be characterized by the boundary state
which contains information about one-point correlation functions of the bulk
operators $V_\alpha=e^{2\alpha\phi}$ on the disk
with the boundary condition labeled by $s$ \cite{FZZbrane}
\be
\langle B_s| = \int_{0}^{\infty}dP\, U(\alpha(P);s)=
\int_{0}^{\infty}dP\, \cos (2\pi P s) \Psi(P)\langle P|,
\label{fzzbst}
\ee 
where $\alpha(P)=Q/2-iP$ and 
\be
\Psi(P)= \(\pi\mul\gamma(b^2)\)^{-{iP\over b}}
{\Gamma(1+2iPb)\Gamma\left(1+{2iP/b}\right)\over i\pi P}.
\label{Psiss}
\ee

The Dirichlet boundary conditions, which are also called ZZ branes, appear 
as non-equivalent quantizations of Lobachevskiy geometry on the world sheet and
thus describe branes living at $\phi=\infty$ .
Similarly to the previous case, they are characterized by a boundary state.
It turns out that at the quantum level there is a two-parameter family of consistent
boundary conditions. They are referred as $(m,n)$ ZZ branes where $m$ and $n$
run over positive integers. The corresponding boundary states are \cite{ZZ}
\be
\langle B_{m,n}|= 2 C \int_{0}^{\infty}dP\,
\sinh (2\pi n P b)\sinh ({2\pi m P/ b})\Psi(P)\langle P|
\label{ZZbst}
\ee
with $C$ being some numerical constant.
 
Other correlation functions of bulk and boundary operators on the disk,
which are not given in \Ref{fzzbst} and \Ref{ZZbst},
were found in \cite{TeschnerBnd,Hos,TeschPons}.

\subsection{Complex curve of minimal string theories}

At first sight the two types of boundary conditions seem to be completely independent.
But it is easy to check that they satisfy the following property \cite{MARTINEC,Hos,Pons}
\be
\langle B_{m,n}|=C\left[\langle B_{s(m,n)}|-\langle B_{s(m,-n)}|\right],
\qquad {\rm where} \qquad
s(m,n)=i\left({m\over b}+n b\right).
\label{difFZZZZ}
\ee
Note that the two values of the parameter $s$, $s(m,n)$ and $s(m,-n)$, correspond
to the same value of the boundary cosmological constant $\mub$. This hints
that the relation \Ref{difFZZZZ} realizes a monodromy property 
of the FZZ boundary state continued analytically to complex values of $\mub$ \cite{TeschnerRim}. 
And indeed, in the case of $c<1$ string theories, 
a nice geometric interpretation for \Ref{difFZZZZ} was found in terms
of a complex curve, which also provided a unified description for different kinds of 
D-branes \cite{Seib}.

The curve comes from two sectors of the theory. Its first origin is the ground ring
of closed string vertex operators. In this way the curve encompasses an information 
about the closed string background. The second origin of the complex curve is
the disk partition function with Neumann boundary condition on the Liouville field 
which is interpreted as the amplitude of an open string ending on the FZZ brane. 
Let us introduce two variables,
$x$ and $y$, related to the boundary cosmological constant and the FZZ partition 
function, respectively
\be
x={\mub} \sim \cosh(\pi b s),
\qquad
y={\p \ZFZZ\over \p {\mub}} \sim \cosh(\pi s/ b).
\label{xycurve}
\ee
Then, considered as complex variables, they satisfy some algebraic relation, $F(x,y)=0$,
which represents an equation of the complex curve embedded into $\Cb^2$.
From the definition \Ref{xycurve}, it follows that the FZZ partition function
is given by a line integral on the curve of the holomorphic differential
$ydx$, whereas the property \Ref{difFZZZZ}
ensures that the disk partition function of the $(m,n)$ ZZ brane 
is evaluated by a similar integral along a closed contour $\gamma_{m,n}$
going from and returning to the point $(x_{m,n}=x(s(m,n)),y_{m,n}=y(s(m,n)))$ 
$$
\ZFZZ(\mub)=\int\limits^{\mub}_{\CP} ydx,  
\qquad
\ZZZ_{m,n}=C\oint\limits_{\gamma_{m,n}} ydx. 
$$
It was shown \cite{Seib} that the points $(x_{m,n},y_{m,n})$ coincide with the singularities 
of the complex curve where it touches itself and forms a ``pinched cycle", 
so that the contours $\gamma_{m,n}$ are non-contractible. Thus, in the $c<1$ case 
each ZZ brane is associated with a singularity of the complex curve which provides 
a unified description for both ground ring and FZZ branes.

\subsection{The $c=1$ limit of the $c<1$ curve}

To generalize the above picture to the case of the $c=1$ string theory, let us
take the limit $b\to 1$ in equations \Ref{xycurve} defining the complex curve.
It is known that this limit is singular and requires the following renormalization of 
couplings
\be
\muc=\mathop{\lim}\limits_{b\to 1} \[\pi(1-b^2)\mul\], 
\quad
\mubc=\mathop{\lim}\limits_{b\to 1} \[\pi(1-b^2)\mub\].
\label{renorm}
\ee
However, it is easy to see that it is not enough because even 
after the renormalization, for example, the FZZ partition function remains singular.
However, the singular term is ``non-universal" from the point of view of the $c=1$ 
theory since it is polynomial in $\mub$. Thus, this term should be subtracted
so that we define the renormalized one-point disk correlation function with FZZ boundary
condition as follows
\be
w(s)\equiv \mathop{\lim}\limits_{b\to 1}
\({ \p_{\mub} \ZFZZ\over \pi(1-b^2)}  + {4 \,  \over \pi}\ZD
 \,\mub \),
\label{renw}
\ee
where $\ZD$ is the disk partition function in the $c=1$ CFT. Then the simple calculation
leads to the following result \cite{SAcurve}
\be
\mubc=\sqrt{\muc}\cosh(\pi s) ,
\qquad
w(s)=-D\sqrt{\muc}\, \pi s \sinh(\pi s), 
\label{rencurve}
\ee
where $D$ is some constant. This gives a parametric representation of the limiting curve
coming from the FZZ partition function.
Contrary to the previous case, all singularities of this curve collapse
just to two points. Indeed, the $c=1$ limit of \Ref{difFZZZZ} reads
\be
s(m,n)=i(m+n),  \qquad
\(\mubc(m,n),w(m,n)\)= \((-1)^{m+n}\sqrt{\muc},0\).
\label{rensing}
\ee
As a result, the curve is degenerate and does not allow to make unambiguous identification
between the singularities and the ZZ branes. Also, as we will see, it differs 
from the curve associated with the ground ring of the $c=1$ string theory.
To understand these issues, to construct a non-degenerate curve and to 
find some non-perturbative effects, we turn now to the matrix model formulation
of the two-dimensional string theory.

\section{Complex curve of the perturbed MQM}

\subsection{Complex curve from the profile of the Fermi sea}

The matrix model in question is the double scaled matrix quantum mechanics.  
More precisely, we are interested in its singlet sector where it can be reduced 
to a system a free fermions in the inverted oscillator potential.
At the quasiclassical level the system of free fermions can be completely 
characterized by the shape of the Fermi sea in the phase space formed
by the matrix eigenvalue $x$, playing the role of the fermion coordinate,  
and its conjugated momentum $p$.

The ground state, corresponding to the simplest linear dilaton background on
the string side, is described by the Fermi sea of the hyperbolic shape,
$\hf(x^2-p^2)=\mu$, where $\mu$ is the Fermi level. Let us parameterize it 
in the way similar to \Ref{rencurve}
\be
x(\tau)=\sqrt{2\mu}\cosh(\tau) ,
\qquad
y(\tau)=\sqrt{2\mu} \sinh(\tau). 
\label{mqmcurve}
\ee
Then we can continue the parameter $\tau$ to the complex plane and view the equation
for the profile of the Fermi sea as an equation for the complex curve associated with 
the given solution of MQM.
This curve differs from the curve \Ref{rencurve} obtained from the FZZ partition function.
In fact, it is this curve that describes the ground ring in the $c=1$ case.
It is easy to see if one introduces the so called light-cone coordinates in the phase space
\be
\xpm=\frac{x\pm p }{ 2}.
\label{xpm}
\ee
In terms of these coordinates the equation for the curve takes the simple form $\xp\xm=\mu$
and coincides with the equation found by Witten \cite{Witgr} provided one identifies $\xpm$
with the generators of the ground ring.

But the complex curve \Ref{mqmcurve} is even more degenerate than the one from \Ref{rencurve}
since it covers itself infinitely many times.
We expect that the degeneracy will disappear after we perturb the theory by some relevant 
operators. The simplest way to do this is to introduce a tachyon potential
characterized by couplings $\lambda_n$. Thus, we are going to consider the theory with 
the following action  
\be
S_{c=1}=\int_{\Sigma}{d^2 z\over 4\pi}\left[(\partial X)^2 +(\partial\phi)^2
+2\hat \CR\phi+\mul  e^{2\phi}+\sum\limits_{n\ge 1} \lambda_n \, e^{(2-{n\over R})\phi}
\cos \(\frc{nX}{ R}\)\right].
\label{pertL}
\ee
In MQM such a closed string background is described by a time-dependent Fermi sea.
Its profile is determined as a consistent solution of the following two equations \cite{AKK}
\be
\xp\xm=
\frc{1}{R} \sum\limits_{k\ge 1} k\lambda_k \, \xpm^{ k/R}  +\mu  +
\frc{1}{R}\sum\limits_{k\ge 1} v_{k}(\mu,\lambda)\,  \xpm^{-k/R},
\label{prof}
\ee
where the coefficients $v_{k}$ contain information about one-point correlation functions
and are fixed by the compatibility condition.
The solution is given in the parametric form \cite{AKK}
\be
\xpm(\tau) = e^{-{1\over 2R}\p^2_{\mu}\CF_0} \,
e^{\pm \tau} \(1+ \sum\limits_{k\ge 1}a_k(\mu,\lambda) \, e^{\mp {k\tau/ R}} \),
\label{solxpm}
\ee
where the coefficients $a_k$ can be found explicitly and $\CF_0$ is the free energy on the sphere. 

As we learned above, the profile of the Fermi sea defines the MQM complex curve. 
Thus, the solution \Ref{solxpm} is what we are looking for! If $\tau\in \Cb$, 
$(\xp(\tau),\xm(\tau))$ defines an embedding of the complex curve into $\Cb^2$ 
and the parameter $\tau$ is its uniformization parameter. It is easy to check that
for non-vanishing couplings the curve is not degenerate anymore.
It possesses singularities given by a one-parameter set of points 
$\tau_n=i\theta_n\in i\Rb$ satisfying $\xpm( \tau_n)=\xpm(- \tau_n)$. 
Thus, $\theta_n$ can be found from the following algebraic equation
\be
\sin\theta_n=\sum\limits_{k\ge 1} a_k\, \sin\(\(\frc{k}{R}-1\)\theta_n\),
\label{thetan}
\ee
where the solutions are ordered in such way that $\theta_n=\pi n +O(\lambda)$.
For a generic value of the parameter $R$ all of them are different, whereas
for rational $R$ only a finite set survives. A more detailed analysis of 
the complex curve can be found in \cite{SAcurve}.

\subsection{Non-perturbative effects from the MQM complex curve}

Similarly to the case of $c<1$, we expect that the complex curve constructed here 
contains information about D-branes and therefore
it is able to describe some of the non-perturbative effects.
We will be interested in a particular kind of such effects which are
given by non-perturbative corrections to the closed string partition function.
This quantity is represented in the matrix model by the free energy and is a sum
of the perturbative and non-perturbative parts
\be
\CF =  \CF_{\rm pert}  + \CF_{\rm non-pert}
=  \sum c_n \, g_s^{2n-2} + \sum a_n\,e^{-d_n/g_s}
\label{expfree}
\ee
The first one is the series in the string coupling and gives closed string amplitudes,
whereas the second part has an exponential form and, as it was shown for critical string 
theories \cite{Polch} and latter in the leading order for non-critical strings \cite{KAK,KK},
describes amplitudes of open strings attached to D-instantons.

The analysis of the fermion system in the presence of arbitrary tachyon perturbation
introduced in \Ref{pertL} allows to calculate the non-perturbative coefficients 
$d_n$ and $a_n$ explicitly. We refer to \cite{AK} for details of the calculation.
The result reads
\beq
d_n &=& i\int_{\gamma_n} \xm d\xp,
\label{dndn}
\\
a_n &\sim & \[  \sin^2{\theta_n \over R} \,
\(\left.{\p\xp\over\p \tau}\right|_{-i\theta_n} \!\! \left.{\p\xm\over\p \tau}\right|_{i\theta_n }-
\left.{\p\xp\over\p \tau}\right|_{i\theta_n}\left. \!\! {\p\xm\over\p \tau}\right|_{-i\theta_n }\)
 \]^{-1/2},
\label{anan}
\eeq
where $\theta_n$ were defined to parameterize the singularities of the complex curve
in \Ref{thetan} and the contours $\gamma_n$ are the images of the intervals 
$(i\theta_n,-i\theta_n)$ under the map \Ref{solxpm}. 
One observes that each non-perturbative correction
is associated with one of the singularities. This points toward
the relation between the singularities and the localized ZZ branes which was found before
for $c<1$ string theories.

\section{CFT interpretation of the MQM results}

To establish connection of the previous results with the CFT description,
one should know correlation functions in the perturbed theory \Ref{pertL}.
Unfortunately, it seems to be a hopeless problem at the present moment. 
Therefore, our method is to expand the matrix model results in the couplings and
to compare the first two terms of the expansion.
For example, if only the first coupling $\lambda_1$ is non-vanishing, these 
terms should correspond to the correlation functions of the cosmological constant 
and the Sine-Liouville operators, respectively, which can be calculated taking the $b= 1$
limit of the results of \cite{FZZbrane,ZZ}
\be
Z(\laml)=\mathop{\lim}\limits_{b\to 1}\(
\ZD\int d\mu \langl V_b \rangl +
\laml \langl \cos\(\frc{X}{R}\)\rangl_{\rm Dir} \langl V_{b-{1\over 2R}} \rangl 
+ \ \dots \ {\mbox{multi-point} \atop \mbox{correlators}}\ \dots 
\).
\label{expan}
\ee
If these terms are the same in MQM and CFT, we claim that the coincidence remains to be true
for the whole series so that the two formulations give the same result for
the quantity under consideration at any values of the couplings.
Then the other terms of the expansion 
provide a matrix model prediction for multi-point correlation functions. 

Following this approach and taking into account 
the relation between the matrix model and CFT couplings, one can show \cite{SAcurve}
that the MQM curve is exactly the same as a CFT curve provided the latter is defined through
the FZZ partition function as follows
\be
x(s)=\sqrt{2\mu}\,\cosh\(\pi s\) \sim \mubc, 
\qquad
p(s)=-{C \over 2i }
\[\p_{x} \ZFZZ\(s+i\) -\p_{x} \ZFZZ\(s-i\)\].
\label{identc}
\ee
This result is quite natural because it is well known that
$\p_{x} \ZFZZ$ is proportional to the matrix model resolvent, whereas 
the momentum $p$ measures the ``width" of the Fermi sea, {\it i.e.}, 
the density of eigenvalues. Thus, the identification \Ref{identc}
is nothing else but the standard relation between the density and the resolvent.
Note also that the first equation in \Ref{identc} allows to find 
the relation between the uniformization parameter
$\tau$ and the CFT parameter $s$. They coincide (up to the factor $\pi$)
only in the non-perturbed case, whereas in general one is a complicated function of the other. 

As an example of a matrix model prediction, we give the two-point correlation function 
on the disk of the Sine-Liouville operator with FZZ boundary conditions on the Liouville 
and Dirichlet conditions on the matter field, which follows from the
second order term in the expansion of \Ref{identc} \cite{SAcurve}
\be
\langl \(\int d^2\sigma e^{\(2-{1\over R}\)\phi}\cos\frc{X}{R}\)^2 \rangl
= -{\pi\Gamma^2\(1-\oR\)\laml^2 \muc^{\oR-1}\over 2^{5/4}\sqrt{R}\,\Gamma^2\(\oR\)}
\( s \coth\(\pi s\)
+ {\sinh\(\({2 \over R}-1\)\pi s\)\over \sin {2\pi \over R}\sinh\(\pi s\)}
\right).
\label{twopcor}
\ee

To complete the identification between the MQM and CFT structures, 
it remains to show that the singularities of the complex curve labeled by $\theta_n$
from \Ref{thetan} are in one to one correspondence with ZZ branes.
Since each singularity of the curve gives rise to a non-perturbative correction to 
the free energy \Ref{expfree}, the coefficients $a_n$ and $d_n$ should have an
interpretation in terms of correlation functions of open strings ending on the ZZ branes.
Indeed, it is easy to check \cite{KAK,SAmn,SAcurve}
(again in the first two orders in the coupling constants) that
the leading correction $d_n$ coincides with the partition function on the disk with
$(n,1)$ ZZ boundary conditions
\be
-d_n=i\int_{\gamma_n} p\, dx=\ZZZ_{n,1}. 
\label{zzcor}
\ee
This means that the $n$th singularity of the complex curve
is associated with the $(n,1)$ ZZ brane and only this set of branes survives in the $c=1$ 
limit \cite{SAmn}.

However, the interpretation of the subleading non-perturbative contribution $a_n$
is still lacking. It should be related to the annulus amplitude 
$\ZZZ_{\rm annulus}(n,1;n,1)$ between two $(n,1)$ branes, but the known expression 
for this quantity diverges \cite{annul}. Nevertheless, it can be obtained from 
two-point correlators of a Gaussian field arising after bosonization of the chiral fermions 
\cite{AK}. But the connection of this field to either ZZ or FZZ branes is not clear to us.

\section{Discussion}

We presented the construction of the complex curve of the $c=1$ string theory.
The formulation in terms of matrix quantum mechanics allowed to find the curve
for the theory perturbed by a tachyon potential. 
In contrast to the non-perturbed case, the resulting curve is not degenerate 
and its singularities are associated with the set of $(n,1)$ ZZ branes.
As in the $c<1$ case, the disk partition functions of these branes 
are given by contour integrals on the curve passing through the singularities.

An important distinction with $c<1$ string theories is that
in the $c=1$ case the complex curve of the ground ring coincides with the curve defined 
by the density of matrix eigenvalues, whereas for $c<1$ it is associated with
the matrix model resolvent. The usual relation between the density and the resolvent allows
to obtain one curve from the other. One can show that this transformation does not affect
the singularities and their relation to D-branes \cite{SAcurve}.

The knowledge of the results for the perturbed theory allows to predict many correlation
functions with FZZ and ZZ boundary conditions. It is enough to expand the results in 
the couplings and make a careful identification of all quantities. We gave here just 
one example --- the two-point correlator of the Sine--Liouville operator.
However, up to now not all found quantities have a known counterpart in the CFT formulation. 
An important problem which remains open is to find such an interpretation for the subleading 
non-perturbative correction. Thus, the relation between MQM and the continuum formulation
beyond the leading order remains still mysterious.

\section*{Acknowledgement}
It is a pleasure to thank V. Kazakov, I. Kostov and D. Kutasov for collaboration.
Also the author is grateful to the organizers of the RTN Workshop 
``The quantum structure of spacetime and the geometric nature of fundamental interactions"
in Kolymbari for the kind hospitality.


\begin{thebibliography}{100}


\bibitem{KLEBANOV}
I. Klebanov, 
{\it String theory in two dimensions,}
\hepth{9108019}.

\bibitem{Moore}
P. Ginsparg and G. Moore,
{\it Lectures on 2D gravity and 2D string theory,}
\hepth{9304011}.

\bibitem{SAthese}
S. Alexandrov,
{\it Matrix quantum mechanics and two-dimensional string theory
in non-trivial backgrounds,} PhD thesis,
\hepth{0311273}.

\bibitem{McGreevyKB}
{J.~McGreevy and H.~Verlinde,
{\it Strings from tachyons: The $c = 1$ matrix reloaded,}
\jhep{0312}{2003}{054},
\hepth{0304224}.
}

\bibitem{MARTINEC}
{E.J. Martinec, {\it The Annular Report on Non-Critical String Theory,}
\hepth{0305148}.
}

\bibitem{KlebanovKM}
{I.R.~Klebanov, J.~Maldacena, and N.~Seiberg,
{\it D-brane decay in two-dimensional string theory,}
\jhep{0307}{2003}{045},
\hepth{0305159}.
}

\bibitem{Seib}
{N. Seiberg and D. Shih, {\it Branes, Rings and Matrix Models in
Minimal (Super)string Theory,}
\jhep{0402}{2004}{021},
\hepth{0312170}.
}

\bibitem{KK}
{V.A. Kazakov and I.K. Kostov,
{\it Instantons in Non-Critical strings from the Two-Matrix Model,}
\hepth{0403152}.
}

\bibitem{SAmn}
{S. Alexandrov,
{\it $(m,n)$ ZZ branes and the $c=1$ matrix model,}
\pl{604}{2004}{115},
\hepth{0310135}.
}

\bibitem{SAcurve}
{S. Alexandrov,
{\it D-branes and complex curves in $c=1$ string theory,}
\jhep{0405}{2004}{025},
\hepth{0403116}.
}

\bibitem{FZZbrane}
V.~Fateev, A.B.~Zamolodchikov and A.B.~Zamolodchikov,
{\it Boundary Liouville field theory. I: Boundary state and boundary  two-point
function,}
\hepth{0001012}.

\bibitem{ZZ}
A.B.~Zamolodchikov and A.B.~Zamolodchikov,
{\it Liouville field theory on a pseudosphere,}
\hepth{0101152}.

\bibitem{TeschnerBnd}
J.~Teschner,
{\it Remarks on Liouville theory with boundary,}
\hepth{0009138}.

\bibitem{Hos}
K.~Hosomichi,
{\it Bulk-Boundary Propagator in Liouville Theory on a Disk,}
\jhep{0111}{2001}{044},
\hepth{0108093}.

\bibitem{TeschPons}
B.~Ponsot and J.~Teschner,
{\it Boundary Liouville field theory: Boundary three point function,}
\np{622}{2002}{309},
\hepth{0110244}.

\bibitem{Pons}
B.~Ponsot,
{\it Liouville Theory on the Pseudosphere: Bulk-Boundary Structure Constant,}
\pl{588}{2004}{105},
\hepth{0309211}.

\bibitem{TeschnerRim}
J.~Teschner,
{\it On boundary perturbations in Liouville theory and brane dynamics in
non-critical string theories,}
\jhep{0404}{2004}{023},
\hepth{0308140}.

\bibitem{Witgr}
{E. Witten, {\it Ground Ring of two dimensional string theory,}
\np{373}{1992}{187}, \hepth{9108004}.
}

\bibitem{AKK}
{S.Yu. Alexandrov, V.A. Kazakov, and I.K. Kostov,
{\it Time-dependent backgrounds of 2D string theory,}
\np{640}{2002}{119}, 
\hepth{0205079}.
}

\bibitem{Polch}
J.~Polchinski,
{\it Combinatorics Of Boundaries In String Theory,}
\physrev{50}{1994}{6041}, 
\hepth{9407031}.

\bibitem{KAK}
{S.Yu. Alexandrov, V.A. Kazakov, and D. Kutasov,
{\it Non-Perturbative Effects in Matrix Models and D-branes,}
\jhep{0309}{2003}{057},
\hepth{0306177}.
}

\bibitem{AK}
{S.Yu. Alexandrov and I.K. Kostov,
{\it Time-dependent backgrounds of 2D string theory: Non-perturbative effects,}
\hepth{0412223}.
}

\bibitem{annul}
D.~Kutasov, K.~Okuyama, J.~Park, N.~Seiberg and D.~Shih,
{\it Annulus amplitudes and ZZ branes in minimal string theory,}
\jhep{0408}{2004}{026},
\hepth{0406030}.





\end{thebibliography}
\end{document}